\documentclass[runningheads]{llncs}
\usepackage[T1]{fontenc}
\usepackage{amssymb, amsmath, bm, latexsym, comment}
\usepackage{xcolor}
\usepackage{graphicx}
\usepackage{setspace}
\usepackage{verbatim}
\usepackage{array}
\usepackage{cite}
\usepackage{booktabs}
\usepackage{hyperref}
\usepackage[capitalize]{cleveref}
\usepackage{arydshln}
\usepackage{multirow}
\usepackage{dsfont}
\usepackage{mathtools}
\usepackage{multirow}
\usepackage{hyphenat}
\usepackage{diagbox}
\usepackage{tikz}
\usepackage{enumitem}
\usepackage[misc,geometry]{ifsym}
\usepackage[linesnumbered,ruled,vlined]{algorithm2e}
\SetAlCapNameFnt{\small}
\usepackage{nicefrac}
\usepackage{physics}
\usepackage{subcaption}
\usepackage{environ}
\usepackage{orcidlink}

\graphicspath{{./figs/}}

\DeclareMathOperator*{\argmin}{arg\,min}

\crefname{algocf}{Alg.}{Algs.}
\Crefname{algocf}{Algorithm}{Algorithms}

\DeclarePairedDelimiterX{\kldivx}[2]{\big[}{\big]}{%
  #1\;\delimsize\|\;#2%
}

\newcolumntype{L}[1]{>{\raggedright\let\newline\\\arraybackslash\hspace{0pt}}m{#1}}
\newcolumntype{C}[1]{>{\centering\let\newline\\\arraybackslash\hspace{0pt}}m{#1}}
\newcolumntype{R}[1]{>{\raggedleft\let\newline\\\arraybackslash\hspace{0pt}}m{#1}}

\newlength{\myl}
\let\origequation=\equation
\let\origendequation=\endequation
\RenewEnviron{equation}{
  \settowidth{\myl}{$\BODY$}                       
  \origequation
  \ifdimcomp{\the\linewidth}{>}{\the\myl}
  {\ensuremath{\BODY}}                             
  {\resizebox{\linewidth}{!}{\ensuremath{\BODY}}}  
  \origendequation
}

\newcommand{\magenta}[1]{\textcolor{magenta}{#1}}
\newcommand{\cyan}[1]{\textcolor{cyan}{#1}}

\begin{document}

\title{Unsupervised Accelerated MRI Reconstruction via Ground-Truth-Free Flow Matching}

\titlerunning{Unsupervised Accelerated MRI Reconstruction}

\author{Xinzhe Luo\inst{1}\orcidlink{0000-0003-2822-1633} \and Yingzhen Li\inst{2}\orcidlink{0000-0001-6938-2375} \and Chen Qin\inst{1}\orcidlink{0000-0003-3417-3092}}

\authorrunning{X. Luo, Y. Li, and C. Qin}   

\tocauthor{}

\institute{
Department of Electrical and Electronic Engineering \& I-X, Imperial College London, London, UK \\
\email{\{x.luo,c.qin15\}@imperial.ac.uk}
\and 
Department of Computing, Imperial College London, London, UK \\
\email{yingzhen.li@imperial.ac.uk}
}

\maketitle

\begin{abstract}
Accelerated magnetic resonance imaging involves reconstructing fully sampled images from undersampled k-space measurements.
Current state-of-the-art approaches have mainly focused on either end-to-end supervised training inspired by compressed sensing formulations, or posterior sampling methods built on modern generative models.
However, their efficacy heavily relies on large datasets of fully sampled images, which may not always be available in practice.
To address this issue, we propose an unsupervised MRI reconstruction method based on ground-truth-free flow matching (GTF\textsuperscript{2}M).
Particularly, the GTF\textsuperscript{2}M learns a prior denoising process of fully sampled ground-truth images using only undersampled data.
Based on that, an efficient cyclic reconstruction algorithm is further proposed to perform forward and backward integration in the dual space of image-space signal and k-space measurement.
We compared our method with state-of-the-art learning-based baselines on the fastMRI database of both single-coil knee and multi-coil brain MRIs.
The results show that our proposed unsupervised method can significantly outperform existing unsupervised approaches, and achieve performance comparable to most supervised end-to-end and prior learning baselines trained on fully sampled MRI, while offering greater efficiency than the compared generative model-based approaches.

\end{abstract}
\section{Introduction}
\label{sec:intro}
Magnetic resonance imaging (MRI) provides a non-invasive diagnostic modality that generates high-resolution depictions of anatomical structures and physiological processes within the human body.
However, the sequential acquisition of measurements in the frequency domain (i.e., k-space) results in prolonged MRI acquisition time, imposing substantial demands on patients and making the process costly and less accessible.
One common strategy to accelerate this process is to undersample the data in k-space and then reconstruct the fully-sampled signals, which naturally leads to an ill-posed inverse problem.
Conventionally, the reconstruction can be achieved by compressed sensing (CS)-based optimisation procedures, via solving a variational problem with sparsity constraints in some transform domain \cite{journal/tit/candes2006, journal/mrm/lustig2007, journal/spm/lustig2008, journal/mrm/liang2009}.
Inspired by that, learning-based methods have been proposed to incorporate network-induced prior and unroll the optimisation steps of the variational problem into an end-to-end training framework \cite{journal/tmi/schlemper2017, journal/mrm/hammernik2018, journal/tmi/aggarwal2018, journal/tpami/yang2018, journal/tmi/qin2019}.
This can be typically achieved by adopting the half quadratic splitting technique that alternates between a de-aliasing step using a regularisation network and a proximal mapping update for measurement consistency.

More recently, accelerated MRI has been approached through the lens of Bayesian inference.
To solve the general inverse problems, these methods build upon generative models that learn the prior distribution of the ground-truth signals (i.e., fully-sampled images), followed by adapting the reverse sampling process to generate samples from the posterior distribution \cite{conference/iclr/song2021, conference/iclr/song2022, conference/iclr/song2023, conference/iclr/chung2023, conference/iclr/wang2023, journal/tmlr/pokle2024}.
For instance, Song et al. \cite{conference/iclr/song2022}, Chung et al. \cite{journal/media/chung2022} and Wang et al. \cite{conference/iclr/wang2023} proposed to incorporate data consistency projections into the reverse sampling process of an unconditional score-based diffusion models \cite{conference/iclr/song2021}.
Further, DPS \cite{conference/iclr/chung2023} and $\Pi$GDM \cite{conference/iclr/song2023} devised strategies to approximate the intractable posterior score for posterior sampling of the ground truth.
However, both end-to-end and prior learning approaches require large datasets of fully-sampled images, which can be hard to access in real-world scenarios.
In addition, the high number of neural function evaluations (NFEs) (typically 100-2000) of the sampling process makes diffusion model-based MRI reconstruction less appealing in practice. 

To overcome the above limitation, unsupervised learning methods for accelerated MRI have also recently raised research attention and been explored in previous work. 
Particularly, the robust equivariant imaging (REI) framework \cite{conference/cvpr/chen2022} proposed to exploit the group invariance property of the signal space, and employed Stein's Risk Estimator (SURE) \cite{journal/ans/stein1981} to enforce measurement consistency.
The ENsemble SURE (ENSURE) framework \cite{journal/tmi/aggarwal2022} further proposed an unbiased estimate of the true mean squared error (MSE) between the prediction and the ground truth, which was then leveraged to train a model-based neural network for unsupervised MRI reconstruction.

In this work, we propose an unsupervised framework for MRI reconstruction, alleviating the demand for fully sampled ground-truth data.
Inspired by ENSURE, we introduce a ground-truth-free learning framework to learn a flow-based denoiser network of fully sampled MR images, based on an induced forward model over the dual-space conditional vector fields.
Moreover, a novel cyclic reconstruction algorithm is proposed that decouples data consistency updates from the training of the denoiser.
The contributions of this work are summarised as:
\begin{itemize}
  \item We introduce a ground-truth-free flow matching framework that learns a prior denoising process for fully sampled MR images using only undersampled k-space measurements.
  \item We propose an efficient cyclic reconstruction algorithm that forward integrates the induced k-space flow to obtain the hidden variable generating the aliased image, and then backward integrates the image-space flow to obtain the target image in a decoupled continuous de-aliasing process.
  \item We evaluated our method on both single-coil and multi-coil accelerated MRI reconstruction. 
  Empirical results demonstrate that the proposed method outperformed existing unsupervised approaches and achieved comparable performance to most state-of-the-art end-to-end and prior learning baselines trained with fully sampled images, whereas with significantly lower NFEs than other generative model-based approaches.
\end{itemize}

\section{Methodology}
\label{sec:method}
\begin{figure}[!t]
    \centering
    \includegraphics[width=\textwidth]{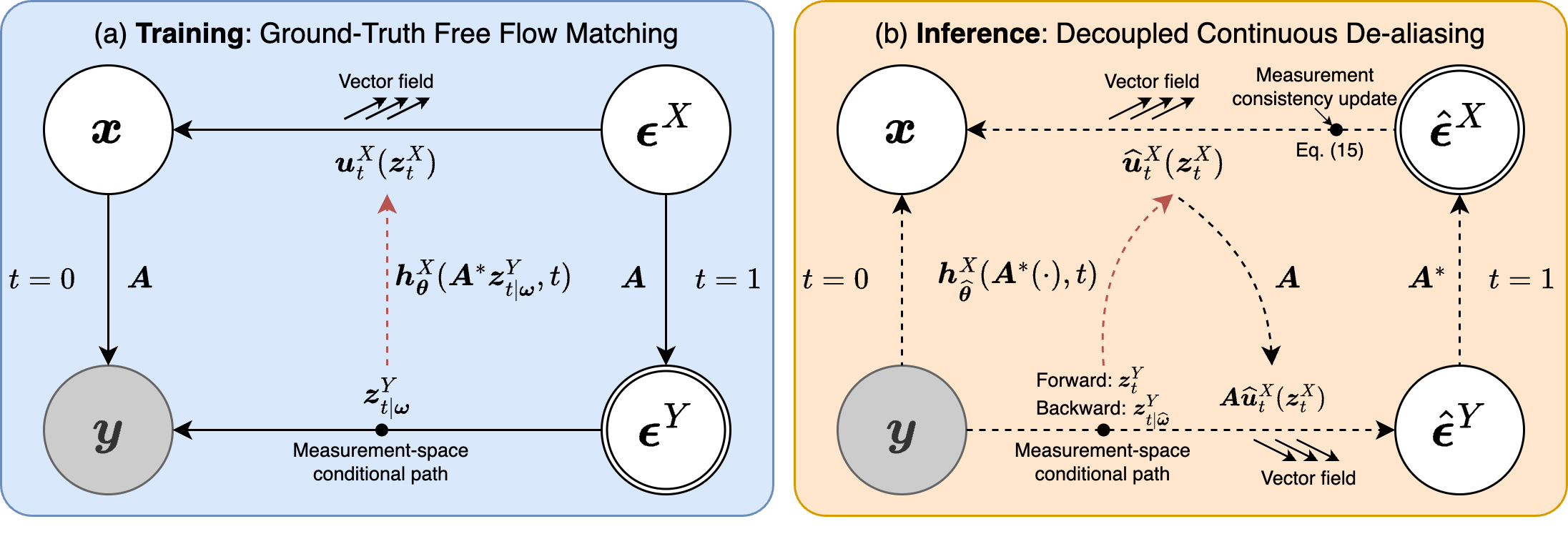}
    \caption{Illustration of the proposed unsupervised reconstruction framework.
    \textbf{(a)} Graphical representation of the training framework via ground-truth-free flow matching. 
    \textbf{(b)} Commutative diagram of the inference steps using cyclic reconstruction to form a decoupled continuous de-aliasing process.
    Random variables are in circles, deterministic variables are in double circles, and observed variables are shaded.
    Solid and dashed arrows indicate the generation and inference steps, respectively.}
    \label{fig:diagram}
\end{figure}

In MRI reconstruction, we denote $\bm{x}\in\mathds{C}^D$ as the image-space \emph{signal}, and $\bm{y}\in\mathds{C}^{d}$ as the undersampled k-space \emph{measurement}.
The forward model is defined as $\bm{y}=\bm{A}\bm{x}+\bm{e}$, where noise $\bm{e}$ follows a complex Gaussian distribution $\mathcal{CN}(\bm{0},2\sigma^2\bm{I}_{d})$, and $\bm{A}\in\mathds{C}^{d\times D}$ is the forward operator combining the undersampling mask $\bm{M}=\mqty[\bm{I}_d\mid\bm{0}]\bm{T}\in\mathds{R}^{d\times D}$ for some permutation matrix $\bm{T}\in\mathds{R}^{D\times D}$, the Fourier transform matrix $\bm{F}\in\mathds{C}^{D\times D}$, and for the multi-coil case the sensitivity map $\bm{S}\in\mathds{C}^{D\times D}$.
To reconstruct the fully-sampled signal, we propose a two-step strategy in our proposed method:
\textbf{(a)} a denoising process of the signal is learnt using flow matching to model the prior of the fully sampled images, with undersampled measurement only; this is achieved by formulating an induced forward model on the dual-space conditional vector fields and optimising a ground-truth-free unbiased estimator of the conditional flow matching objective;
\textbf{(b)} a decoupled continuous de-aliasing process with a novel cyclic integration algorithm is then performed to reconstruct the signal based on the learnt denoiser;
this involves forward integration to the hidden variable generating the aliased undersampled image, and backward measurement-consistent integration to the target image.
\cref{fig:diagram} presents an overview of the proposed framework.

\subsection{Dual-Space Conditional Vector Fields}
Flow matching aims to learn a continuous normalising flow that progressively transforms a simple base distribution $p_1$ into a more complex target distribution $p_0$ of the ground-truth signals \cite{conference/iclr/liu2023, conference/iclr/lipman2023, conference/icml/esser2024}.
The transformation is defined by a diffeomorphic mapping $\phi_t:[0,1]\times\mathds{R}^D\rightarrow\mathds{R}^D$, which can be obtained with an ordinary differential equation (ODE) given a time-dependent vector field $\bm{u}_t:[0,1]\times\mathds{R}^D\rightarrow\mathds{R}^D$, 
\begin{equation}
  \bm{z}_t\triangleq\phi_t(\bm{z}_1),\quad \dd\bm{z}_t = \bm{u}_t(\bm{z}_t)\,\dd t, \quad\forall\,\bm{z}_1\sim p_1,
\end{equation}
with the boundary condition $\phi_1=\operatorname{id}$.
The base distribution $p_1$ usually follows a standard Gaussian, and thus $\phi_t(\cdot)$ can be viewed as a continuous denoising function with $\bm{u}_t(\cdot)$ as its infinitesimal residual.

The (marginal) vector field $\bm{u}_t(\cdot)$ generating the probability path $p_t=[\phi_t]_{\sharp}p_1$ (the push-forward distribution of $p_1$ by $\phi_t$) can be constructed via the conditional path and vector field.
Let $\bm{z}_{t\mid\bm{\omega}}^X\triangleq\psi_t^X(\bm{\omega}^X)=a_t\bm{x}+b_t\bm{\epsilon}^X$ be the conditional path in the ground-truth signal space $X$, where $\bm{x}\in\mathds{C}^D\sim p_0$, $\bm{\epsilon}^X\sim p_1=\mathcal{CN}(\bm{0},2\bm{I}_D)$, and $\bm{\omega}^X\triangleq(\bm{x},\bm{\epsilon}^X)\sim p_{\bm{\omega}}^X\in\mathcal{C}(p_0,p_1)$ which is the coupling of $p_0$ and $p_1$.
We define the conditional vector field $\bm{u}_{t\mid\bm{\omega}}^X(\cdot)$
\begin{equation}
  \bm{u}_{t\mid\bm{\omega}}^X(\cdot)\triangleq\frac{\dd \psi_t^X(\cdot)}{\dd t}=a_t'\bm{x}+b_t'\bm{\epsilon}^X,\quad \forall\,t\in[0,1],
\end{equation}
where $a_t'=\frac{\dd a_t}{\dd t}$ and $b_t'=\frac{\dd b_t}{\dd t}$, such that $a_0=b_1=1$ and $a_1=b_0=0$.
One common instantiation is the independent coupling $p_{\bm{\omega}}^X=p_0\times p_1$ and linear interpolation $\psi_t^X(\bm{\omega}^X)=(1-t)\bm{x}+t\bm{\epsilon}^X$ with $p_{t}(\cdot\mid\bm{\omega}^X)=\delta_{(1-t)\bm{x}+t\bm{\epsilon}^X}(\cdot)$, where $\delta$ is the Dirac delta function. 
Then, the marginal vector field $\bm{u}_t$ that generates $p_t$ can be constructed by 
\begin{equation}
  \bm{u}_t^X(\bm{z}_t) \triangleq\mathbb{E}_{\bm{\omega}^X\sim p_t(\cdot\mid\bm{z}_t)}\left[\bm{u}_{t\mid\bm{\omega}}^X(\bm{z}_t)\right]= \mathbb{E}_{\bm{\omega}^X\sim p_0\times p_1}\left[\bm{u}_{t\mid\bm{\omega}}^X(\bm{z}_t)\frac{p_t(\bm{z}_t\mid\bm{\omega}^X)}{p_t(\bm{z}_t)}\right],
\end{equation}
which satisfies the continuity equation \cite{book/villani2009}.

Given the above probability path defined with $\bm{u}_t^X(\cdot)$, flow matching learns the generative mapping between $p_1$ and $p_0$ by regressing the vector field with a predictor network $\bm{v_{\theta}}^X(\cdot)$ using the conditional flow matching objective
\begin{equation}\label{eq:signal_cfm}
  \mathcal{L}_{\text{CFM}}(\bm{\theta}) = \mathbb{E}_{t\sim p_T,\bm{\omega}^X\sim p_0\times p_1,\bm{z}_{t\mid\bm{\omega}}^X\sim p_t(\cdot\mid\bm{\omega}^X)}\norm{\bm{v_{\theta}}^X(\bm{z}_{t\mid\bm{\omega}}^X,t)-\bm{u}_{t\mid\bm{\omega}}^X(\bm{z}_{t\mid\bm{\omega}}^X)}_2^2.
\end{equation}
However, the scarcity of the ground-truth (i.e., fully-sampled) signal $\bm{x}$ in MRI makes the signal-space conditional path inaccessible. 
To address this, we instead define a measurement-space ($Y$) conditional path $\bm{z}_{t\mid\bm{\omega}}^Y\triangleq\psi_t^Y(\bm{y},\bm{\epsilon}^Y)=a_t\bm{y}+b_t\bm{\epsilon}^Y$, where $\bm{y}=\bm{Ax}+\bm{e}$ and $\bm{\epsilon}^Y\triangleq\bm{A\epsilon}^X$.
Therefore, the signal- and measurement-space conditional vector fields take the form of
\begin{equation}\label{eq:conditional_vec}
  \bm{u}_{t\mid\bm{\omega}}^X(\bm{z}_{t\mid\bm{\omega}}^X) \triangleq a_t'\bm{x} + b_t'\bm{\epsilon}^X,\ 
  \text{and} \ \bm{u}_{t\mid\bm{\omega}}^Y(\bm{z}_{t\mid\bm{\omega}}^Y) \triangleq a_t'\bm{y} + b_t'\bm{\epsilon}^Y.
\end{equation}
Substituting $\bm{y}=\bm{Ax}+\bm{e}$ and $\bm{\epsilon}^Y=\bm{A\epsilon}^X$ into \cref{eq:conditional_vec} yields 
\begin{equation}\label{eq:vector_forward}
  \bm{u}_{t\mid\bm{\omega}}^Y(\bm{z}_{t\mid\bm{\omega}}^Y) = \bm{A}\bm{u}_{t\mid\bm{\omega}}^X(\bm{z}_{t\mid\bm{\omega}}^X) + a_t'\bm{e}.
\end{equation}
This implies that the original signal-measurement forward model $\bm{y}=\bm{Ax}+\bm{e}$ induces another forward model over the dual-space conditional vector fields, which we can exploit to optimise the signal-space conditional flow matching objective in an unsupervised ground-truth-free manner.

\subsection{Ground-Truth-Free Flow Matching (GTF\textsuperscript{2}M)}
In contrast to \cref{eq:signal_cfm} which aims to predict the signal-space conditional vector field based on the \emph{signal-space} conditional path that requires the knowledge of ground-truth signals,
here we propose an unsupervised objective to regress the signal-space conditional vector field from the \emph{measurement-space} conditional path with a predictor network $\bm{h_{\theta}}^X(\cdot)$, namely,
\begin{equation}
  \mathcal{L}_{\text{GTF\textsuperscript{2}M}}(\bm{\theta}) \triangleq \mathbb{E}_{p_T,p_0\times p_1,p_{t}^X(\bm{z}_{t\mid\bm{\omega}}^X\mid\bm{\omega}^X),p_t(\bm{z}_{t\mid\bm{\omega}}^Y\mid\bm{z}_{t\mid\bm{\omega}}^X)}\norm{\bm{h_{\theta}}^X(\bm{z}_{t\mid\bm{\omega}}^Y,t)-\bm{u}_{t\mid\bm{\omega}}^X(\bm{z}_{t\mid\bm{\omega}}^X)}_2^2,
\end{equation}
where $p_t(\bm{z}_{t\mid\bm{\omega}}^Y\mid\bm{z}_{t\mid\bm{\omega}}^X)$ is induced from $\bm{z}_{t\mid\bm{\omega}}^Y=\bm{Az}_{t\mid\bm{\omega}}^X+a_t\bm{e}$.
In addition, similar to Theorem 2 in \cite{conference/iclr/lipman2023}, we can prove that
\begin{equation}\label{eq:sf2m}
  \mathcal{L}_{\text{GTF\textsuperscript{2}M}}(\bm{\theta}) = \mathbb{E}_{p_T,p_t^X(\bm{z}_t^X),p_t(\bm{z}_t^Y\mid\bm{z}_t^X)}\norm{\bm{h_{\theta}}^X(\bm{z}_t^Y,t)-\bm{u}_t^X(\bm{z}_t^X)}_2^2 + \operatorname{const.},
\end{equation}
which means we are essentially predicting the signal-space marginal vector field.

Meanwhile, noting from \cref{eq:conditional_vec} that $\bm{z}_{t\mid\bm{\omega}}^Y=a_t\bm{y}+b_t\bm{\epsilon}^Y=\frac{a_t}{a_t'}\bm{u}_{t\mid\bm{\omega}}^Y(\bm{z}_{t\mid\bm{\omega}}^Y)-b_t'\left(\frac{a_t}{a_t'}-\frac{b_t}{b_t'}\right)\bm{\epsilon}^Y$, we can write $\bm{h_{\theta}}^X(\bm{z}_{t\mid\bm{\omega}}^Y,t)=\bm{h_{\theta}}^X\!\left(\bm{u}_{t\mid\bm{\omega}}^Y(\bm{z}_{t\mid\bm{\omega}}^Y),t\right)$.
Taking into account the randomness in the forward operator $\bm{A}_s$ indexed by some random variable $s\sim p(s)$ (i.e., a random permutation matrix $\bm{T}_s$ encoding the positions of k-space acquisition), the ENSURE framework \cite{journal/tmi/aggarwal2022} implies that the GTF\textsuperscript{2}M objective has an unbiased estimate up to a constant, which does not require the fully sampled signal, namely,
\begin{equation}
  \begin{aligned}
    &\mathbb{E}_{s,t,\bm{\omega}^X,\bm{z}_{t\mid\bm{\omega},s}^X,\bm{z}_{t\mid\bm{\omega},s}^Y\mid\bm{z}_{t\mid\bm{\omega},s}^X}\norm{\bm{h_{\theta}}^X(\bm{z}_{t\mid\bm{\omega},s}^Y,t)-\bm{u}_{t\mid\bm{\omega},s}^X(\bm{z}_{t\mid\bm{\omega},s}^X)}_2^2 \\
    =\ &\mathbb{E}_{s,t,\bm{\omega}^X,\bm{z}_{t\mid\bm{\omega},s}^X,\bm{z}_{t\mid\bm{\omega},s}^Y\mid\bm{z}_{t\mid\bm{\omega},s}^X}\norm{\bm{R}_s\!\left[\bm{h_{\theta}}^X\!\left(\bm{u}_{t\mid\bm{\omega},s}^Y(\bm{z}_{t\mid\bm{\omega},s}^Y),t\right)-\bm{u}_{t\mid\bm{\omega},s}^X(\bm{z}_{t\mid\bm{\omega},s}^X)\right]}_2^2 \\
    =\ & \mathbb{E}_{s, t,\bm{\omega}^Y,\bm{z}_{t\mid\bm{\omega},s}^Y}
    \left[
    \norm{\bm{R}_s\!\left[\bm{h_{\theta}}^X(\bm{\mu}_{{t\mid\bm{\omega}},s}^X,t)-\widehat{\bm{u}}_{{t\mid\bm{\omega}},{s},\text{ML}}^X\right]}_2^2 + 2\nabla_{\bm{\mu}_{{t\mid\bm{\omega}},s}^X}\cdot\bm{R}_s^*\bm{R}_s\bm{h_{\theta}}^X(\bm{\mu}_{{t\mid\bm{\omega}},s}^X,t)
    \right] + \operatorname{const.}
  \end{aligned}
\end{equation}
where $\bm{\mu}_{{t\mid\bm{\omega}},{s}}^X\triangleq\bm{A}_s^*\bm{C}_t^{-1}\bm{u}_{{t\mid\bm{\omega}},{s}}^Y(\bm{z}_{t\mid\bm{\omega},s}^Y)$ is a sufficient statistic for $\bm{u}_{t\mid\bm{\omega},s}^X(\bm{z}_{t\mid\bm{\omega},s}^X)$ with $\bm{C}_t=(a_t'\sigma)^2\bm{I}_d$, and $\widehat{\bm{u}}_{{t\mid\bm{\omega}},{s},\text{ML}}^X\triangleq (\bm{A}_s^*\bm{C}_t^{-1}\bm{A}_s)^{\dag}\bm{A}_s^*\bm{C}_t^{-1}\bm{u}_{{t\mid\bm{\omega}},{s}}^Y(\bm{z}_{t\mid\bm{\omega},s}^Y)$ is the maximum likelihood solution of the forward model in \cref{eq:vector_forward}.
The projection operator is defined as $\bm{R}_s\triangleq\bm{W}\bm{P}_s$ with $\bm{P}_s=\bm{A}_s^{\dag}\bm{A}_s$ and $\bm{W}\triangleq\mathbb{E}_s[\bm{P}_s]^{-\nicefrac{1}{2}}$.
For single-coil reconstruction where $\bm{A}_s=\bm{M}_s\bm{F}$, $\bm{R}_s$ has a closed-form expression as
$\bm{R}_s = \bm{F}^*\bm{T}_s^{\intercal}\operatorname{diag}(p_1^{-0.5},\dots,p_d
^{-0.5},0,\dots,0)\bm{T}_s\bm{F}$, where $p_i$ denotes the probability that the $i$-th k-space measurement is acquired.

By ignoring the constant and rewriting $\bm{h_{\theta}}^X(\bm{\mu}_{{t\mid\bm{\omega}},s}^X,t)=\bm{h_{\theta}}^X(\bm{A}_s^*\bm{z}_{t\mid\bm{\omega},s}^Y,t)$, we derive the GTF\textsuperscript{2}M objective $\mathcal{L}_{\text{GTF\textsuperscript{2}M}}(\bm{\theta})$ equivalent to 
\begin{equation}\label{eq:objective}
  \mathbb{E}_{s,t,\bm{\omega}^Y,\bm{z}_{t\mid\bm{\omega},s}^Y}
  \left[
    \norm{\bm{R}_s\!\left[\bm{h_{\theta}}^X(\bm{A}_s^*\bm{z}_{t\mid\bm{\omega},s}^Y,t)-\widehat{\bm{u}}_{t,{s},\text{ML}}^X\right]}_2^2 +
    2a_t'a_t\sigma^2\nabla_{\bm{A}_s^*\bm{z}_{t\mid\bm{\omega},s}^Y}\cdot \bm{R}_s^*\bm{R}_s\bm{h_{\theta}}^X(\bm{A}_s^*\bm{z}_{t\mid\bm{\omega},s}^Y,t)
  \right],
\end{equation}
which is essentially \emph{ground-truth-free}.
That is, the objective learns the ground-truth signal-space marginal vector field without fully sampled MRI data.

In practice, the divergence term can be computed using the Hutchinson trace estimator \cite{journal/CSSC/hutchinson1989}, namely,
\begin{equation}
  \nabla_{\bm{A}_s^*\bm{z}_{t\mid\bm{\omega},s}^Y}\cdot\bm{R}_s^*\bm{R}_s\bm{h_{\theta}}^X(\bm{A}_s^*\bm{z}_{t\mid\bm{\omega},s}^Y,t) = \mathbb{E}_{\bm{b}\sim\mathcal{N}(\bm{0},\bm{R}_s^*\bm{R}_s)}\left[\bm{b}^{\intercal}\nabla_{\bm{A}_s^*\bm{z}_{t\mid\bm{\omega},s}^Y}\bm{h_{\theta}}^X(\bm{A}_s^*\bm{z}_{t\mid\bm{\omega},s}^Y,t)\bm{b}\right],
\end{equation}
where the expectation can be approximated by Monte-Carlo sampling.

\subsection{Reconstruction as Decoupled Continuous De-aliasing}
To reconstruct the fully sampled image $\bm{x}$ from the observed undersampled k-space measurement $\bm{y}$, we further propose a cyclic reconstruction algorithm based on the vector field learnt by the GTF\textsuperscript{2}M objective.
Given the signal-space vector field $\bm{u}_t^X(\cdot)$ and a random noise $\bm{z}_1^X=\bm{\epsilon}^X\sim p_1$, we can denoise it via \emph{backward} integration of the ODE, namely
\begin{equation}\label{eq:backward_int}
  \bm{z}_t^X=\bm{z}_1^X+\int_1^t\bm{u}_t^X(\bm{z}_t^X)\,\dd t,
\end{equation}
and particularly at $t=0$ we have $\bm{z}_0^X= \bm{\epsilon}^X + \int_1^0\bm{u}_t^X(\bm{z}_t^X)\,\dd t\sim p_0$.
The marginal vector field $\bm{u}_t^X(\bm{z}_t^X)$ is estimated from $\bm{h_{\theta}}^X(\bm{A}^*\bm{z}_{t\mid\bm{\omega}}^Y)$ learnt by the GTF\textsuperscript{2}M objective, where $\bm{z}_{t\mid\bm{\omega}}^Y=a_t\bm{y}+b_t\bm{A\epsilon}^X$ given the observed measurement $\bm{y}$.

However, instead of directly using a random $\bm{\epsilon}^X$, we propose starting from the aliased initial point $\bm{A}^*\bm{A}\widehat{\bm\epsilon}^X$ corresponding to the aliased image $\bm{A}^*\bm{y}$, where $\widehat{\bm\epsilon}^X$ is the unknown hidden variable generating the ground truth $\bm{x}$  via the learnt vector field $\widehat{\bm{u}}_t^X(\bm{z}_t^X)$. 
This is analogous to unrolling-based learning methods that take as input the aliased image $\bm{A}^*\bm{y}$.
Specifically, applying the MRI forward model to both sides of the \emph{forward} ODE $\bm{z}_t^X=\bm{x}+\int_0^t\bm{u}_t^X(\bm{z}_t^X)\,\dd t$, we can obtain
\begin{equation}\label{eq:forward_ode}
  \bm{z}_t^Y\triangleq\bm{Az}_t^X + \bm{e} = \bm{y} + \int_0^t\bm{Au}_t^X(\bm{z}_t^X)\,\dd t.
\end{equation}
Therefore, we propose to estimate a better initial point $\bm{z}_1^X\triangleq\bm{A}^*\widehat{\bm{\epsilon}}^Y$ by
\begin{equation}
  \widehat{\bm{\epsilon}}^Y\triangleq\bm{A}\widehat{\bm\epsilon}^X\approx\bm{A}\widehat{\bm\epsilon}^X+\bm{e} = \bm{y} + \int_0^1\bm{A}\widehat{\bm{u}}_t^X(\bm{z}_t^X)\,\dd t,
\end{equation}
where the approximation comes from the fact that $\norm{\bm{A}\widehat{\bm\epsilon}^X}$ follows a Chi distribution whose expectation is proportional to the standard deviation of the \emph{i.i.d.} Gaussian variables of $\widehat{\bm\epsilon}^X$, and we can always increase its variance by design.
Since we do not have access to the true trajectory $\bm{z}_t^X$ of the ground truth $\bm{x}$, the above estimation is intractable.
However, if the estimated signal-space flow is straight enough \cite{conference/iclr/liu2023}, we can integrate the following ODE as an approximation to \cref{eq:forward_ode}:
\begin{equation}
  \bm{z}_t^Y = \bm{y} + \int_0^t\bm{A}\bm{h_{\theta}}^X(\bm{A}^*\bm{z}_t^Y)\,\dd t,
\end{equation}
which gives an estimate $\widehat{\bm{\epsilon}}^Y=\bm{z}_1^Y$ and then we set $\bm{z}_1^X \triangleq \bm{A}^* \widehat{\bm{\epsilon}}^Y$.

Then, starting from the new initial point $\bm{z}_1^X=\bm{A}^*\widehat{\bm\epsilon}^Y$, we backward integrate \cref{eq:backward_int} by injecting measurement consistency updates, giving rise to a continuous de-aliasing process as the number of discrete Euler steps goes to infinity.
Specifically, given $K$ uniform time points $t=\nicefrac{1}{K},\dots,1$ and , \cref{eq:backward_int} can be approximated by discrete Euler steps
\begin{equation}
  \bm{z}_{t-\nicefrac{1}{K}}^X \leftarrow \bm{z}_{t}^X - \frac{1}{K}\bm{h_{\theta}}^X(\bm{A}^*\bm{z}_{t\mid\bm{\omega}}^Y).
\end{equation}
To align $\bm{Az}_{t}^X$ with $\bm{z}_{t\mid\bm{\omega}}^Y$ for ensuring data consistency, we propose to perform proximal mapping to enforce measurement consistency via the optimisation problem 
\begin{equation}\label{eq:proximal_mapping}
  \bm{z}_t^X \leftarrow \argmin_{\bm{z}}\left\{{\gamma_t}\norm{\bm{Az}-\bm{z}_{t\mid\bm{\omega}}^Y}_2^2+\norm{\bm{z}-\bm{z}_t^X}_2^2\right\},
\end{equation}
where the regularisation coefficient $\gamma_t=\frac{\zeta\sigma_t}{(a_t\sigma)^2}=\frac{\zeta}{b_t^2\sigma^2}$ \cite{coference/cvpr/zhu2023}, with $\sigma_t\triangleq\nicefrac{a_t^2}{b_t^2}$ the signal-to-noise ratio (SNR) at time $t$ and $\zeta$ a hyper-parameter. 
In single-coil setting, a closed-form solution of \cref{eq:proximal_mapping} can be dervied as
\begin{equation}
  \bm{z}_t^X \leftarrow \bm{z}_t^X - \lambda_t\bm{A}^*(\bm{A}\bm{z}_t^X-\bm{z}_{t\mid\bm{\omega}}^Y),
\end{equation}
where the step size $\lambda_t\triangleq\frac{\gamma_t}{1+\gamma_t}=\left(1+\nicefrac{b_t^2\sigma^2}{\zeta}\right)^{-1}$.
Thus, unlike previous learning-based reconstruction methods that unroll conventional optimisation schemes, the proposed strategy decouples data consistency updates from the training of the denoiser network, leading to improved training efficiency and flexibility. 

Furthermore, the forward and backward integration formulates a cyclic path, which represents a commutative diagram for inferring the ground-truth image as shown in \cref{fig:diagram} (b).

For multi-coil reconstruction, coil-wise reconstruction can be similarly performed following the aforementioned steps, yielding $\bm{z}_{0,c}^X$ for $c=1,\dots,C$, followed by sensitivity combined reconstruction \cite{journal/mrm/pruessmann1999}, i.e., $\widehat{\bm{x}}=\left(\sum_c\bm{S}_c^*\bm{S}_c\right)^{-1}\sum_c\bm{S}_c^*\bm{z}_{0,c}^X$, where $\bm{S}_c$ is a diagonal matrix encoding the sensitivity map of the $c$-th coil.
\Cref{alg:main} summarises the proposed reconstruction process.

\begin{algorithm}[t]
  \small
  \caption{Decoupled Continuous Dealiasing via Cyclic Integration}
  \label{alg:main}
  \KwIn{k-space measurement $\bm{y}=(\bm{y}_c)_{c=1}^C$, pretrained flow predictor $\bm{h}_{\widehat{\bm\theta}}^X(\cdot,t)$, forward steps $L$, backward steps $K$, regularisation parameter $\zeta$}
  \KwOut{Reconstructed image $\widehat{\bm{x}}$ of $\bm{y}$}
  ${\bm{z}}_0^Y:=\bm{y}$\;
  \For{$t=0,\dots,\nicefrac{(L-1)}{L}$}{
    ${\bm{z}}_{t+\nicefrac{1}{L}}^Y \leftarrow{\bm{z}}_{t}^Y + \frac{1}{L}\bm{A}\bm{h}_{{\widehat{\bm\theta}}}^X(\bm{A}^*{\bm{z}}_{t}^Y,t)$\tcp*{forward integration}
  }
  $\widehat{\bm{\epsilon}}^Y\triangleq\bm{z}_1^Y$, $\bm{z}_1^X \triangleq \bm{A}^* \widehat{\bm{\epsilon}}^Y$\;
  \For{$t\in\{1, \dots, \nicefrac{1}{K}\}$}{
    $\bm{z}_{t\mid\widehat{\bm\omega}}^Y=a_t\bm{y}+b_t\widehat{\bm\epsilon}^Y$\;
    $\bm{z}_{t-\nicefrac{1}{K}}^X \leftarrow \bm{z}_{t}^X - \frac{1}{K}\bm{h}_{\widehat{\bm\theta}}^X(\bm{A}^*\bm{z}_{t\mid\widehat{\bm\omega}}^Y,t)$\tcp*{backward integration}
    $\bm{z}_t^X \leftarrow \bm{z}_t^X - \lambda_t\bm{A}^*(\bm{A}\bm{z}_t^X-\bm{z}_{t\mid\widehat{\bm\omega}}^Y)$\tcp*{measurement consistency update}
  }
  $\widehat{\bm{x}}=\bm{z}_0^X\text{ or }\left(\sum_{c=1}^C\bm{S}_c^*\bm{S}_c\right)^{-1}\sum_{c=1}^C\bm{S}_c^*\bm{z}_{0,c}^X$\tcp*{single- or multi-coil}
  \Return{$\widehat{\bm{x}}$.}
\end{algorithm}

\section{Experiments and Results}
\label{sec:experiment}
\subsection{Experimental Setups}
\noindent\textbf{Datasets and preprocessing.}
We used the NYU fastMRI Initiative database\footnote{\url{https://fastmri.med.nyu.edu/}} \cite{journal/rai/knoll2020, arxiv/zbontar2018} to evaluate the performance of our model for accelerated MRI reconstruction on both the single-coil knee and multi-coil brain datasets.
A selection of 484/56/44 single-coil proton density (PD) weighted knee MRI volumes without fat suppression and 700/100/200 multi-coil T2 weighted brain MRI volumes were used for training/validation/testing.
Each slice was cropped to the size of $320\times 320$ and normalised to a constant norm.
We simulated Cartesian subsampling masks for each volume following the setup in \cite{arxiv/zbontar2018}, which include  8\% and 4\% fully-sampled low-frequency k-space lines for acceleration factors of 4 and 8, respectively.
The other lines were sampled random-uniformly and equidistantly respectively for knee and brain volumes.
For knee data, we used the provided emulated single-coil (ESC) data as the ground truth, while for brain MRI we used the SENSE reconstruction  \cite{journal/mrm/pruessmann1999}, where a spatially invariant noise distribution was assumed.
The coil sensitivity maps were estimated by ESPIRiT \cite{journal/mrm/uecker2014}, following the implementation of \cite{journal/mrm/hammernik2021}.
Finally, reconstruction performance is measured by calculating the structural similarity index measure (SSIM) and the peak signal-to-noise ratio (PSNR) between prediction and ground truth images.

\noindent\textbf{Implementation Details.}
For the conditional path coefficients, we adopt linear interpolation $a_t=1-t$ and $b_t=t$.
The noise level is considered as $\sigma=0.01$.
We use the ADM (ablated diffusion model) network architecture \cite{conference/nips/dhariwal2021} for the flow predictor $\bm{h_{\theta}}^X(\cdot,t)$, which consists of a U-Net with adaptive group normalisation after each intermediate convolutional block.
The parameters of group normalisation are given by linear projection of the positional embeddings of time points.
In addition, multi-resolution attention and dropout with probability $0.1$ are applied at the lowest three resolutions of the network.
We trained the network using the AdamW optimiser \cite{conference/iclr/loshchilov2019} with a learning rate of $1\times 10^{-4}$ and a weight decay coefficient of $0.1$.
Exponential moving average of the network parameters was performed every 100 training steps with rate $0.99$.
The hyperparameter $\zeta$ is set to 1 by default.
The code was implemented in PyTorch \cite{conference/nips/paszke2019} with training and inference performed on NVIDIA L40S and A100 GPUs.

\noindent\textbf{Compared Baselines.}
Three types of baseline methods were compared with our approach:
(a) supervised end-to-end learning on paired samples of fully sampled images and their corresponding measurements, for which we compared MoDL \cite{journal/tmi/aggarwal2018}, an architecture applicable to both single- and multi-coil reconstruction;
(b) generative model-based methods that learn a prior distribution of fully sampled MRI and incorporate measurements into posterior sampling steps, for which we compared DDNM\textsuperscript{+} \cite{conference/iclr/wang2023}, $\Pi$GDM \cite{conference/iclr/song2023} and FlowPS \cite{journal/tmlr/pokle2024};
in particular, DDNM\textsuperscript{+} used range-null decomposition for measurement consistency, while $\Pi$GDM and FlowPS relied on approximation of the posterior score function;
(c) unsupervised approaches without prior learning using only undersampled k-space data , including REI \cite{conference/cvpr/chen2022} and ENSURE \cite{journal/tmi/aggarwal2022}.
Note that except for MoDL and ENSURE that are directly applicable to MRI reconstruction, the other baselines were only developed for either inverse problems of natural images or single-coil MRI.
We adapt them to both single- and multi-coil MRI reconstruction in our experiments.

\subsection{Single-Coil Knee MRI Reconstruction}

\begin{table}[t]
\small
  \centering
  \caption{Qualitative results of 4$\times$ and 8$\times$ accelerated single-coil MRI reconstruction using various reconstruction methods on the fastMRI knee data with the random uniform undersampling pattern.
  Statistical significant difference ($p<0.05$) between ours and the other methods indicated by two-sided paired $t$-tests was marked with asterisks (*).
  Best results among ground-truth-dependant and -free methods are highlighted in \cyan{cyan} and \magenta{magenta}, respectively.}
  \begin{tabular}{C{2cm}C{2.1cm}C{2.1cm}c@{\hspace{0cm}}C{2.1cm}C{2.1cm}C{1.2cm}}
    \toprule
    \multirow{2}{*}{Method} & \multicolumn{2}{c}{SSIM $\uparrow$} & & \multicolumn{2}{c}{PSNR $\uparrow$} & \multirow{2}{*}{NFEs $\downarrow$} \\
    \cmidrule{2-3}\cmidrule{5-6}
    & $4\times$ & $8\times$ & & $4\times$ & $8\times$ & \\
    \midrule
    Zero-filled & $0.684\pm 0.086$* & $0.556\pm0.106$* & & $27.60\pm2.78$* & $23.92\pm 2.75$* & N/A \\
    \hline
    \multicolumn{7}{c}{(a) Supervised methods using fully sampled images} \\
    \hdashline\noalign{\vskip 0.5ex}
    MoDL \cite{journal/tmi/aggarwal2018} & $0.786\pm0.069$* & \cyan{$0.692\pm0.107$}* & & $30.72\pm3.07$* & \cyan{$28.58\pm2.96$}* & 1 \\
    \hline
    \multicolumn{7}{c}{(b) Prior learning methods using fully sampled images} \\
    \hdashline\noalign{\vskip 0.5ex}
    DDNM\textsuperscript{+} \cite{conference/iclr/wang2023} & \cyan{$0.791\pm 0.076$}* & $0.681\pm0.108$* &  & \cyan{$31.73\pm 3.29$}* & $28.00\pm3.20$ & 100 \\
    $\Pi$GDM \cite{conference/iclr/song2023} & $0.728\pm0.098$* & $0.581\pm0.114$* & & $30.27\pm 3.34$* & $25.83\pm3.13$* & 100 \\
    FlowPS \cite{journal/tmlr/pokle2024} & $0.763\pm0.077$* & $0.631\pm0.101$* & & $30.66\pm2.73$* & $26.30\pm2.55$* & 100 \\
    \midrule
    \multicolumn{7}{c}{(c) Unsupervised methods w/o prior learning} \\
    \hdashline\noalign{\vskip 0.5ex}
    REI \cite{conference/cvpr/chen2022} & $0.740\pm0.087$* & $0.591\pm0.110$* & & $29.96\pm2.87$* & $25.04\pm2.97$* & 1 \\
    ENSURE \cite{journal/tmi/aggarwal2022} & $0.684\pm0.086$* & $0.556\pm0.106$* & & $27.65\pm2.79$* & $23.91\pm2.75$* & 1 \\
    \hline
    \multicolumn{7}{c}{(d) Unsupervised methods w/ prior learning} \\
    \hdashline\noalign{\vskip 0.5ex}
    Ours & $\magenta{0.801\pm0.073}$ & \magenta{$0.688\pm0.095$} & & \magenta{$31.64\pm2.95$} & \magenta{$27.95\pm2.64$} & 20 \\
    \bottomrule
  \end{tabular}
  \label{tab:knee_compare}
\end{table}

\begin{figure}[t]
    \centering
    \includegraphics[width=\textwidth]{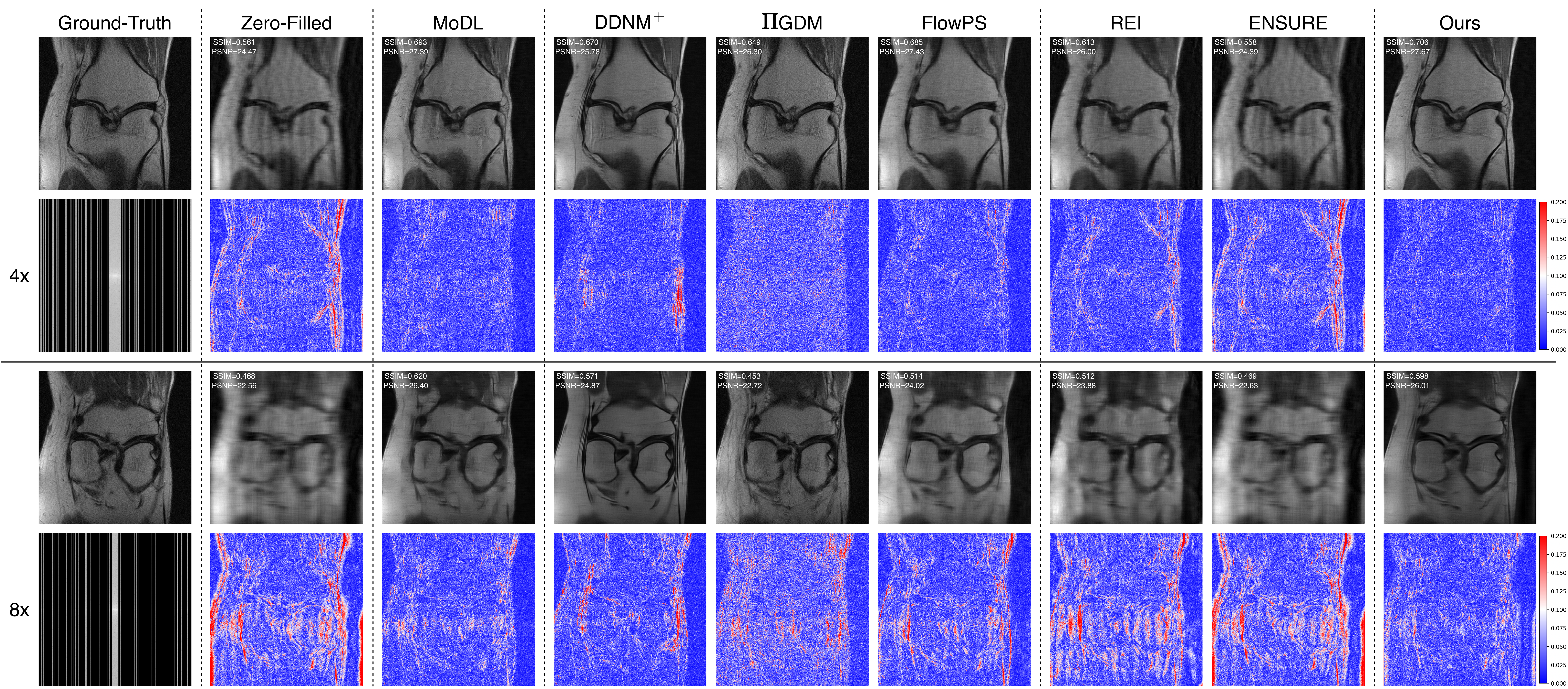}
    \caption{Visualisation of reconstruction on a sample single-coil knee MRI using various compared methods.
    The k-space images are presented in log-scale absolute values.
    The error maps are presented in values relative to the peak intensity in the ground-truth image.
    It can be observed that our approach can achieve comparable performance to that obtained from supervised training using MoDL.}
    \label{fig:knee_vis}
\end{figure}

\noindent\textbf{Comparison study.}
\cref{tab:knee_compare} presents the reconstruction accuracy on the single-coil knee MRI dataset.
Notably, our proposed approach performed significantly better than the unsupervised baselines.
Compared with existing supervised and prior learning approaches that rely on fully sampled images, our method also achieved comparable performance in majority cases.
Moreover, our method achieved superior efficiency than other generative model-based ones during inference, through a substantial reduction in the number of neural function evaluations (NFEs) required.
\cref{fig:knee_vis} visualises reconstruction and its error map on a sample single-coil knee MRI using various compared methods.
In particular, the error map generated by our method is comparable to that produced by MoDL.

\begin{table}[!t]
  \small
  \centering
  \caption{Ablation study on the effect of forward integration for initialisation in the proposed algorithm on single-coil knee MRI reconstruction.
  Statistical significant difference ($p<0.05$) indicated by two-sided paired $t$-tests was marked with asterisks (*).}
  \begin{tabular}{C{2.4cm}|C{2.cm}C{2.cm}c@{\hspace{0cm}}C{2.cm}C{2.cm}C{1.1cm}}
    \toprule
    \multirow{2}{*}{Method} & \multicolumn{2}{c}{SSIM $\uparrow$} & & \multicolumn{2}{c}{PSNR $\uparrow$} & \multirow{2}{*}{NFEs $\downarrow$} \\
    \cmidrule{2-3}\cmidrule{5-6}
    & $4\times$ & $8\times$ & & $4\times$ & $8\times$ & \\
    \midrule
    w/o forward int. & $0.755\pm0.077$* & $0.648\pm0.098$* & & $30.29\pm2.69$* & $27.02\pm2.55$* & 10 \\
    \hdashline\noalign{\vskip 0.5ex}
    w/ forward int. & $0.801\pm0.073$ & $0.688\pm0.095$ & & $31.64\pm2.95$ & $27.95\pm2.64$ & 20 \\
    \bottomrule
  \end{tabular}
  \label{tab:ablation}
\end{table}

\begin{figure}[!t]
    \centering
    \includegraphics[width=\textwidth]{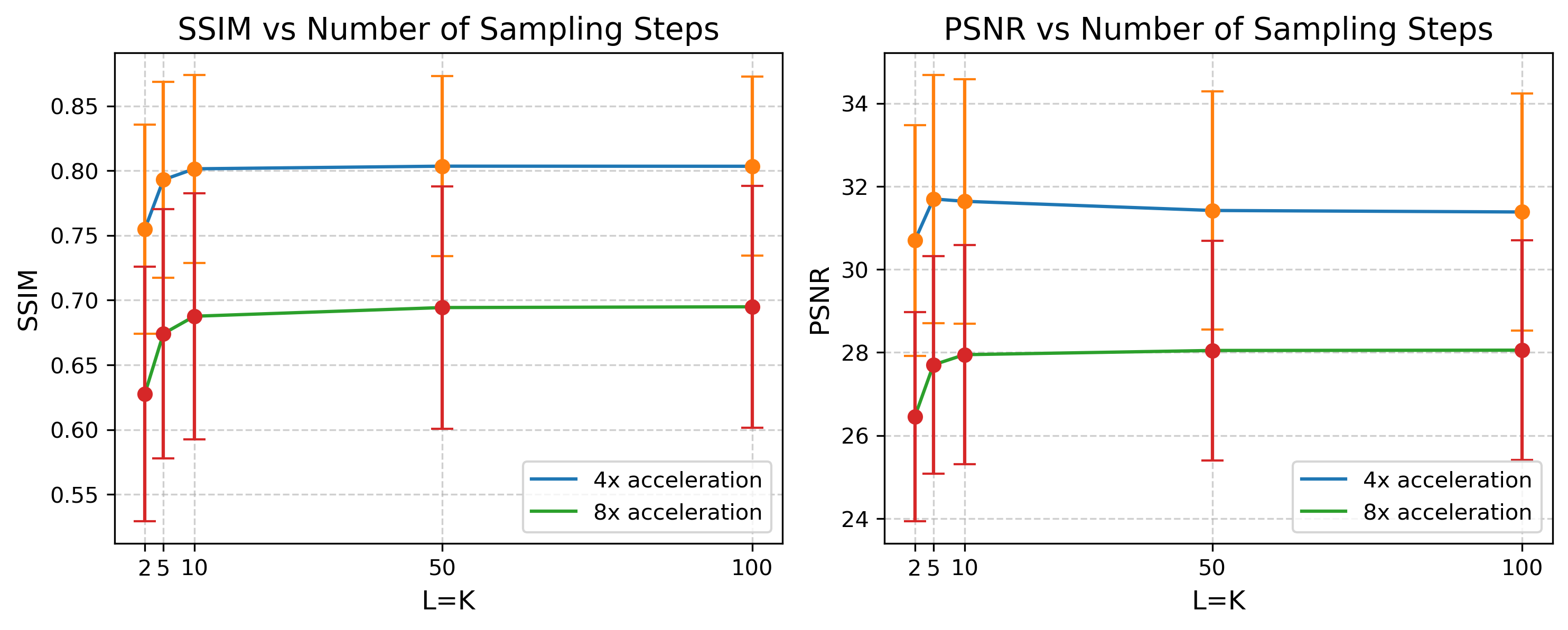}
    \caption{Reconstruction accuracy on single-coil knee MRI as a function of the number of de-aliasing steps in the proposed algorithm.}
    \label{fig:sample_steps}
\end{figure}

\noindent\textbf{Ablation and parameter study.}
\cref{tab:ablation} presents the ablation study on the effect of applying forward integration steps to initialise noise in the proposed decoupled continuous de-aliasing algorithm.
It is evident that the incorporation of forward integration markedly enhances the reconstruction accuracy.
This suggests that a good initialisation, which aligns with the observed measurements, is crucial for the effectiveness of the de-aliasing process.
\cref{fig:sample_steps} presents the reconstruction accuracy as a function of the number of integration steps in the proposed cyclic reconstruction algorithm.
It can be observed that the enhancement in accuracy begins to plateau when the number of forward and backward steps reaches 10.
This justifies our default configuration of $L=K=10$ as an optimal balance between accuracy and efficiency.

\subsection{Multi-Coil Brain MRI Reconstruction}

\begin{table}[!t]
  \small
  \centering
  \caption{Qualitative results of 4$\times$ and 8$\times$ accelerated multi-coil MRI reconstruction using various reconstruction methods on the fastMRI brain data with the uniform equidistant undersampling pattern.
  Statistical significant difference ($p<0.05$) between ours and the other methods indicated by two-sided paired $t$-tests was marked with asterisks.
  Best results among ground-truth-dependant and -free methods are highlighted in \cyan{cyan} and \magenta{magenta}, respectively.}
  \begin{tabular}{C{2.2cm}C{2.1cm}C{2.cm}c@{\hspace{0.cm}}C{2.1cm}C{2.cm}C{1.2cm}}
    \toprule
    \multirow{2}{*}{Method} & \multicolumn{2}{c}{SSIM $\uparrow$} & & \multicolumn{2}{c}{PSNR $\uparrow$} & \multirow{2}{*}{NFEs $\downarrow$} \\
    \cmidrule{2-3}\cmidrule{5-6}
    & $4\times$ & $8\times$ & & $4\times$ & $8\times$ & \\
    \midrule
    Zero-filled & $0.800\pm 0.089$* & $0.716\pm0.117$* & & $27.66\pm3.78$* & $24.10\pm3.97$* & N/A \\
    \hline
    \multicolumn{7}{c}{(a) Supervised methods using fully sampled images} \\
    \hdashline\noalign{\vskip 0.5ex}
    MoDL \cite{journal/tmi/aggarwal2018} & \cyan{$0.948\pm0.044$}* & $0.820\pm0.051$* & & $38.28\pm3.37$* & $30.18\pm3.04$* & 1 \\
    \hline
    \multicolumn{7}{c}{(b) Prior learning methods using fully sampled images} \\
    \hdashline\noalign{\vskip 0.5ex}
    DDNM\textsuperscript{+} \cite{conference/iclr/wang2023} & $0.929\pm 0.045$* & \cyan{$0.887\pm0.048$}* &  & \cyan{$40.61\pm 3.43$}* & \cyan{$34.13\pm2.92$}* & 100 \\
    FlowPS \cite{journal/tmlr/pokle2024} & $0.855\pm0.060$* & $0.748\pm0.069$* & & $33.10\pm2.73$* & $26.56\pm3.50$* & 100 \\
    \midrule
    \multicolumn{7}{c}{(c) Unsupervised methods w/o prior learning} \\
    \hdashline\noalign{\vskip 0.5ex}
    ENSURE \cite{journal/tmi/aggarwal2022} & $0.825\pm0.053$* & $0.739\pm0.108$* & & $31.75\pm3.99$* & $25.56\pm3.50$* & 1 \\
    \hline
    \multicolumn{7}{c}{(d) Unsupervised methods w/ prior learning} \\
    \hdashline\noalign{\vskip 0.5ex}
    Ours & \magenta{$0.920\pm0.060$} & \magenta{$0.859\pm0.054$} & & \magenta{$34.65\pm2.32$} & \magenta{$28.72\pm2.92$} & 20 \\
    \bottomrule
  \end{tabular}
  \label{tab:brain_compare}
\end{table}

\begin{figure}[!t]
    \centering
    \includegraphics[width=\textwidth]{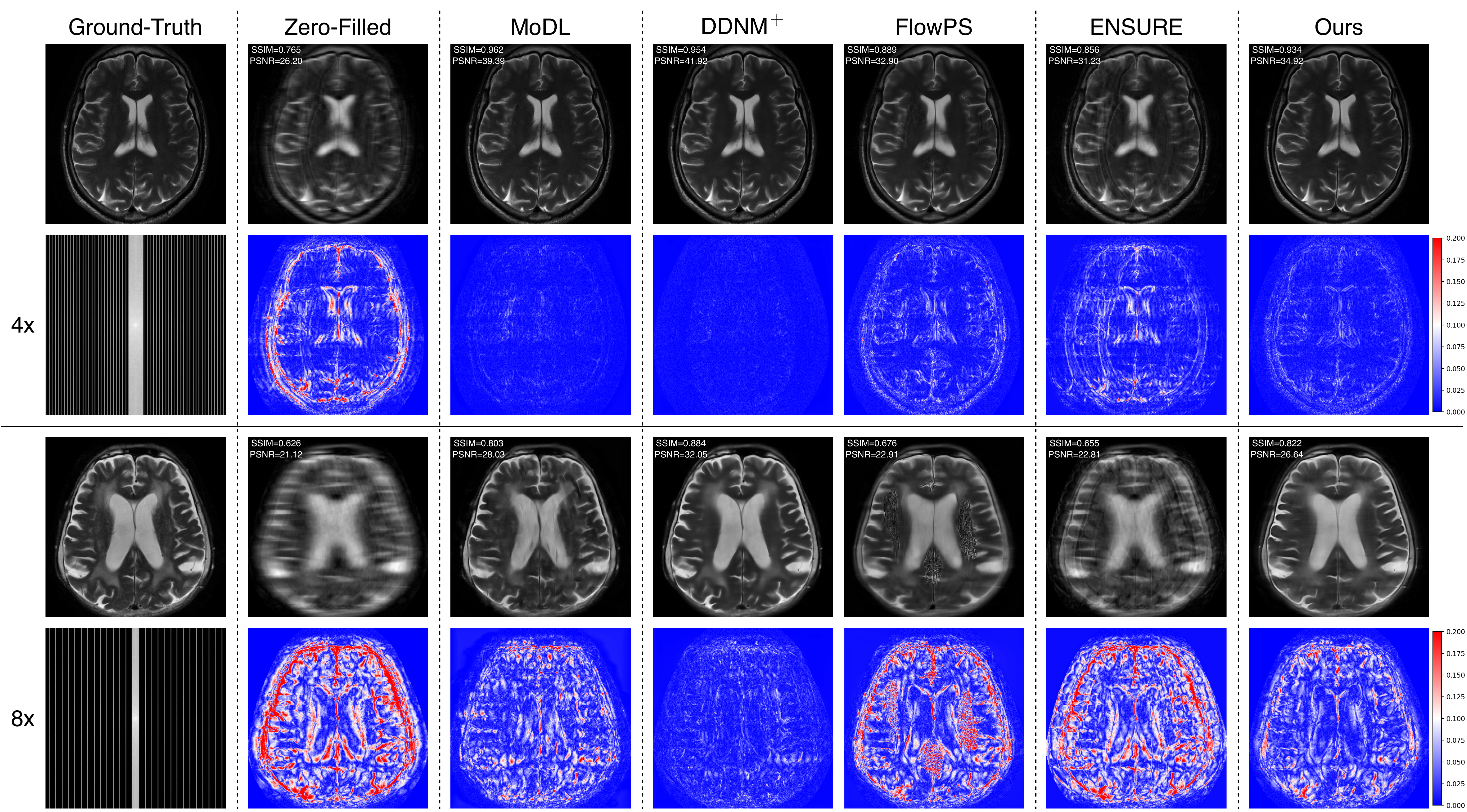}
    \caption{Visualisation of reconstruction on a sample multi-coil brain MRI using various compared methods.
    The k-space images are presented as the log-scale root sum square of the multi-coil undersampled k-space maps.
    The error maps are presented in values relative to the peak intensity in the ground-truth image.}
    \label{fig:brain_vis}
\end{figure}

\textbf{Comparison study.}
\cref{tab:brain_compare} presents the reconstruction accuracy on the multi-coil brain MRI dataset.
Our method performed the best among unsupervised approaches, and better than the flow-based method FlowPS that used fully sampled images for prior learning.
However, there is still a performance gap between ours and the other methods that use fully sampled images.
This could be attributed to the fact that our method is trained by the coil-wise images, which may be less efficient to capture the prior of the coil-combined ground-truth images.
\cref{fig:brain_vis} visualises reconstruction and its error map on a sample multi-coil brain MRI using the compared methods.

\section{Conclusion}
\label{sec:conclusion}
In this work, we have presented an unsupervised framework for accelerated MRI reconstruction based on flow matching.
The proposed model adapts the original flow matching objective into a ground-truth-free variant based on an induced forward model between the dual-space conditional vector fields.
A decoupled continuous de-aliasing process is further proposed through a cyclic reconstruction algorithm, where the forward integration estimates a better initialisation for backward integration, leading to improved reconstruction quality.
The proposed method achieves the best performance among all the unsupervised approaches compared.
However, its reconstruction accuracy may be outperformed by methods that leverage fully sampled images during training, especially in the multi-coil case.
Future work will investigate methodological advancement that can mitigate this performance gap.
For example, we can train the denoiser based on the combined forward operator of multi-coil MRI, which may entail a numerical solver for the projection operator in the proposed GTF\textsuperscript{2}M objective, whereas it might cause instability in the training process.

\subsubsection{Acknowledgement.}
This work was supported by the Engineering and Physical Sciences Research Council (EPSRC) UK grant TrustMRI [EP/X039277/1].

\bibliographystyle{splncs04}
\bibliography{biblio.bib}

\end{document}